\documentclass[conference]{IEEEtran}
\IEEEoverridecommandlockouts
\usepackage{cite}
\usepackage{amsmath,amssymb,amsfonts}
\usepackage{algorithmic}
\usepackage{graphicx}
\usepackage{textcomp}
\def\BibTeX{{\rm B\kern-.05em{\sc i\kern-.025em b}\kern-.08em
    T\kern-.1667em\lower.7ex\hbox{E}\kern-.125emX}}

\IEEEoverridecommandlockouts
\IEEEpubid{\makebox[\columnwidth]{978-1-7281-4490-0/20/\$31.00 \copyright~2020~IEEE }
\hspace{\columnsep}\makebox[\columnwidth]{ }}
 
\newcommand \ignore[1]{}

\usepackage{caption}
\usepackage{subcaption}
\usepackage{siunitx}
\usepackage{multirow}

\usepackage[table]{xcolor}
\usepackage{cases}
\usepackage{array}
\usepackage{color}
\definecolor{pink}{rgb}{1,0,1} 

\begin{document}

\title{Context-Dependent Implicit Authentication for Wearable Device Users
}

\author{%
William~Cheung and Sudip Vhaduri\\
Fordham University, Bronx, NY 10458, USA\\
\{wcheung5, svhaduri\}@fordham.edu\\
}

\ignore{------------------------------------------
\author{\IEEEauthorblockN{1\textsuperscript{st} Given Name Surname}
\IEEEauthorblockA{\textit{dept. name of organization (of Aff.)} \\
\textit{name of organization (of Aff.)}\\
City, Country \\
email address or ORCID}
\and
\IEEEauthorblockN{2\textsuperscript{nd} Given Name Surname}
\IEEEauthorblockA{\textit{dept. name of organization (of Aff.)} \\
\textit{name of organization (of Aff.)}\\
City, Country \\
email address or ORCID}
\and
\IEEEauthorblockN{3\textsuperscript{rd} Given Name Surname}
\IEEEauthorblockA{\textit{dept. name of organization (of Aff.)} \\
\textit{name of organization (of Aff.)}\\
City, Country \\
email address or ORCID}
\and
\IEEEauthorblockN{4\textsuperscript{th} Given Name Surname}
\IEEEauthorblockA{\textit{dept. name of organization (of Aff.)} \\
\textit{name of organization (of Aff.)}\\
City, Country \\
email address or ORCID}
\and
\IEEEauthorblockN{5\textsuperscript{th} Given Name Surname}
\IEEEauthorblockA{\textit{dept. name of organization (of Aff.)} \\
\textit{name of organization (of Aff.)}\\
City, Country \\
email address or ORCID}
\and
\IEEEauthorblockN{6\textsuperscript{th} Given Name Surname}
\IEEEauthorblockA{\textit{dept. name of organization (of Aff.)} \\
\textit{name of organization (of Aff.)}\\
City, Country \\
email address or ORCID}
}
---------------------------------------}

\maketitle

\begin{abstract}
As market wearables are becoming popular with a range of services, including making financial transactions, accessing cars, etc. that they provide based on various private information of a user, security of this information is becoming very important. However, users are often flooded with PINs and passwords in this internet of things (IoT) world. Additionally, hard-biometric, such as facial or finger recognition, based authentications are not adaptable for market wearables due to their limited sensing and computation capabilities. Therefore, it is a time demand to develop a burden-free implicit authentication mechanism for wearables using the less-informative soft-biometric data that are easily obtainable from the market wearables. In this work, we present a context-dependent soft-biometric-based wearable authentication system utilizing the heart rate, gait, and breathing audio signals. From our detailed analysis, we find that a binary support vector machine (SVM) with radial basis function (RBF) kernel can achieve an average accuracy of $0.94 \pm 0.07$, $F_1$ score of $0.93 \pm 0.08$, an equal error rate (EER) of about $0.06$ at a lower confidence threshold of 0.52, which shows the promise of this work. 
\end{abstract}


\section{Introduction}\label{introduction}

\subsection{Motivation}
The interconnected nature of Internet of Things (IoT) have allowed us to remotely collect information or control multitude of physical objects.
\ignore{Various examples of such systems include alarm systems, entertainment devices, vehicles, and smart home devices, to name a few. IoT systems are connected to existing network infrastructure providing new interfaces that integrate the physical world with computer-based systems.} 
Along with the growth of IoT, advancements in smartphones and wearables in their sensing and computational capabilities to a point which enable many new applications and usage scenarios to emerge~\cite{vhaduri2020adherence,vhaduri2020nocturnal,vhaduri2019nocturnal,vhaduri2019towards,al2009load}.
\ignore{ While smartphones are already widely used \cite{seneviratne2017survey},} 
Even with this progress, wearables are still growing in popularity with the arrival of new applications. Some include the ability to identify a user to third party services~\cite{bianchi2016wearable}, protect commercial customer information (i.e., passwords, credit card information)~\cite{nguyen2017smartwatches}, manage financial payments\ignore{~\cite{seneviratne2017survey}}, allows access to smartphones and other paired devices\ignore{~\cite{kumar2016authenticating}}, unlock vehicles~\cite{nguyen2017smartwatches}, monitor or track individuals (e.g., child or elderly monitoring or fall detection), and assess an individual's health and fitness. According to a recent market report, a 72.7\% increase in wearable shipments and an associated increase in sales revenue of 78.1\% are predicted from 2016 to 2022~\cite{wearable_shipments_revenue}. 

However, wearables also raise new challenges, especially in terms of security. The main accuracy and reliability concern is that imposters with unauthorized access can steal information from other sensitive IoT objects, which poses a significant risk~\cite{zeng2017wearia}.
\ignore{Unauthorized users could also steal data on the wearables itself, e.g., many applications and services provided by a wearable depend on sensor and user data stored on the device to grant access.} 
Furthermore, an intentional device sharing between target and non-target users might lead to inaccurate and faulty assessment since healthcare providers and researchers are increasingly relying on wearables to monitor their patients or study participants remotely. Therefore, there is an imperative need for a robust and accurate authentication mechanism specifically for wearable device users.

Existing wearable devices either have no authentication systems or authentication mechanisms that are often knowledge-based regular PIN locks or pattern locks~\cite{nguyen2017smartwatches} \cite{guerar2020circlepin}, which suffer from scalability issues~\cite{unar2014review}. 
\ignore{With an increasing reliance on protected devices, a user can be overwhelmed with passwords or PIN requests to obtain access to various data and services. Knowledge-based approaches also require user interactions with the display, which may either be inconvenient to certain class of users or even completely absent in many wearables~\cite{unar2014review,zeng2017wearia}.}
Additionally, many times, users opt to completely disable security mechanisms out of convenience, as the design hinders the implementation of security itself.

\subsection{Related Work}

\subsubsection{Wearable Constraints}
Wearable device user authentication is a relatively new field of research compared to other mobile authentication~\cite{bianchi2016wearable,blasco2016survey,cornelius2012wears,unar2014review}.
The limited display sizes of wearables add another constraint that limits the choices of authentication mechanisms~\cite{bianchi2016wearable,vhaduri2017wearable}. But as technology advances companies such as Samsung, Fit-bit, Apple, Garmin, and Embrace can provide lower level granularity in data. More biometrics are available as more sensors are being added such as microphone, electrocardiograms (ECG), and GPS but there still hold accuracy concerns. Researchers have found that, although for people over the age of 85 Apple accurately detects atrial fibrillation at a rate of 96\%, for people under 55 it only correctly diagnoses atrial fibrillation 19.6\% of the time~\cite{Apple_ECG}. \ignore{However, given better ECG technology, implicit authentication models will have a better performance.} Another group of researchers~\cite{yan2019towards} developed designed wrist strapped ECG reader and developed an authentication system with an accuracy of 93.5\%, which is limited by the ease of use and the need for user movement. Therefore, an authentication scheme that can utilize data from a multitude of readily available sensors on market wearables could be more realistic to develop a non-stop implicit wearable device user authentication system. 

\subsubsection{Multi-modal Biometric Authentication}
In previous work, combinations of biometrics were used to form multi-modal biometric authentication systems for increased reliability compared to unimodal systems, which often suffer from noisy data, intra-class variations, inter-class similarities, and spoof attacks~\cite{ghayoumi2015review}. For multi-modal authentication systems, researchers have utilized different hard- and soft-biometrics. \rule{0pt}{7ex} However, due to relatively low computational power of wearables, these multi-modal approaches are typically not implemented for implicit and continuous authentication on state-of-the-art wearables.

\subsubsection{Wearable Authentication}
 Researchers recently proposed authentication techniques that are more suitable for wearables, focusing more on approaches based on {\em behavioral biometrics}, such as gait~\cite{al2017unobtrusive,cola2016gait,johnston2015smartwatch}, activity types~\cite{bianchi2016wearable,zeng2017wearia}, gesture~\cite{davidson2016smartwatch}, keystroke dynamics~\cite{acar2018waca} and {\em physiological biometrics}, such as PPG signals~\cite{karimian2017non}. Almost all of these studies are based on project specific generated datasets\ignore{ databases and the accuracy of these techniques is often verified with limited numbers of subjects and over short time periods}. \ignore{All of these user authentication techniques are limited in the scope of use, e.g., gait-based behavioral authentication approaches~\cite{cola2016gait,johnston2015smartwatch} only work during walking.} While other projects have addressed some of the limitations of gait-based approaches by considering different types of gestures~\cite{davidson2016smartwatch} or activities~\cite{bianchi2016wearable,zeng2017wearia}. All of these models are based on movement and thereby, fail to work in the very common human state of being sedentary~\cite{acar2018waca,vhaduri2017wearable}. Authentication approaches using physiological biometric data, such as heart rate and bioimpedance~\cite{cornelius2012wears} require fine-grained samples and sensor readings are easily affected by noise, motion, etc but are constantly available.\ignore{Focusing on one or a particular set of biometrics restricts the usability of a continuous authentication model, but} Depending on a user's context, i.e., phsysical state and availability of biometrics it is possible to build a robust multi-modal authentication process, which is able to continue \ignore{and adjust }changing contexts.

\subsection{Contributions}
The main contribution of this paper is the exploration of a hierarchical non-stop implicit authentication for wearables using less informative coarse-grained soft-biometrics. 
Compared to previous work~\cite{vhaduri2017wearable,vhaduri2019multi}, where we use hybrid-biometrics, such as calorie burn that can be affected by a user's self-reported input, e.g., age, height, and weight, in this work we focus on three different soft-biometrics, i.e., heart rate, gait, and breathing that can be measured without a user's self-reported input and can be easily obtained from the market wearables. 
In this work, we present a multi-biometric-based hierarchical context-driven approach (discussed in Section~\ref{methods}) that works both in {\em sedentary} and {\em non-sedentary} periods. We develop both binary and unary models based on the availability of other people's data in addition to a valid user's data. 
We are able to authenticate a user with an average accuracy of $0.82 \pm 0.08$ \& $F_1 = 0.81 \pm 0.08$  ({\em non-sedentary}, Table~\ref{HG_Results}) and an average accuracy $0.94 \pm 0.07$ \& $F_1 = 0.93 \pm 0.08$ ({\em sedentary}, Table~\ref{HS_Results}), while developing the binary SVM models  (Section~\ref{model_performance}). 
While developing the unary models, we obtain an average accuracy of $0.72 \pm 0.07$ and $F_1 = 0.73 \pm 0.06$ (Sections~\ref{model_performance}).
\section{Approach}\label{approach}

In this paper, we intend to demonstrate the importance and effectiveness of different biometrics to identify wearable device users with the help of different machine learning models. Before we describe the detailed analysis, we first introduce the datasets, pre-processing steps, feature engineering, and methods used in this work. 

\rule{0pt}{4ex} 
\subsection{Datasets}\label{datasets}
In this work, we use the following three different datasets.

\begin{itemize}
    \item Fitbit dataset: We use the heart rate data collected at a rate of one sample per minute using the Fitbit Charge HR device from 10 subjects similar to our previous work\ignore{~\cite{vhaduri2017wearable,vhaduri2017towards,vhaduri2018biometric,vhaduri2019multi,vhaduri2018opportunisticTBD,vhaduri2018hierarchical}}~\cite{vhaduri2016assessing,vhaduri2016cooperative,vhaduri2017discovering,vhaduri2017wearable,vhaduri2017towards,vhaduri2018hierarchical,vhaduri2018biometric,vhaduri2018impact,vhaduri2018opportunistic,vhaduri2018opportunistic,vhaduri2016design,vhaduri2016human,vhaduri2018opportunisticTBD,vhaduri2017design,vhaduri2019multi,chihyou2020estimating,cheung2020context}.
    \item Gait dataset: We use the WISDM dataset~\cite{WISDM}, where gyroscope and accelerometer readings were collected at a rate of one sample in 50 milliseconds using the LG G Watch (running Wear 1.5 operating system). In this work, we use 10 subjects' data. 
    \item Audio dataset: 
    We collect breathing audio clips from 10 subjects with six distinct inhalation breathing events per clip using the Evistr digital voice recorder. The clips are around 5 seconds long. 
\end{itemize}

\subsection{Data Pre-Processing}\label{pre-proc}

Since we are using a real-world datasets, we first need to clean the dataset before using it. Then, we need to segment the continuous stream of biometrics, such as heart rate, gait information, and desired audio events (i.e., breathing). 
\ignore{In addition personal data was added to the data set in order to test robustness and practicality.} 
Finally, we compute and select influential features before constructing authentication models. 

\subsubsection{Data Segmentation}\label{vData}
Since heart rate and gait data were sampled at different frequencies, therefore, we segment the heart rate and gait samples into 10-sample windows\ignore{ in order to synchronize the two time-series data and} obtain stable and rich information\ignore{ before feature generation}. Using a 50\% sliding window, we obtain 800 heart rate windows and 720 gait windows, i.e., instances from each subject.
Unlike the heart rate or gait data, the audio data comes with other types of sounds in addition to desired breathing sounds. Additionally, some clips come with multiple breathing events separated by silence or noisy parts. Therefore, we segment the \ignore{ESC-50 }audio clips to fetch single inhalation breathing events. Thereby, we obtain around six inhalation breathing events per subject. 
Each event is modified in 102 ways mentioned in the next section (Section~\ref{Audio_Data_Augmentation}), we obtain a total of 612 instances from each subject. \ignore{To give a fair representation of performance between}While utilizing the three different biometrics to develop different models discussed in the Methods section, i.e., Section~\ref{methods}, we consider the same 612 instances from each biometric.
\ignore{Per 5 second clip of breathing audio sound,} 
\ignore{The rest of the audio recording that did not satisfy this parameter was not used in this study.}

\subsubsection{Audio Data Augmentation}\label{Audio_Data_Augmentation}
\ignore{Sampling size of some aspect of the data sources were limited and a need to generate simulated estimates of other data was needed in order to facilitated better testing.} 
\ignore{Fingerprints are inherently static but the types of imaging might not be. In order to simulate this, we augmented the original data points. For images changes with various rotations, zooms, and combinations rotation with zooms were used, while for audio pitch and speed adjustments were made.}
Breathing audio could be altered due to a change in contexts, e.g., environments, physical state, or mood. To simulate this and capture the variations, we augment the original audio breathing events using various pitch shifts and speed changes.
\begin{itemize}
    \item Pitch shift: We consider 15 different pitch shifts ranging from -3.5 to 3.5 with 0.5 increments
    \item Speed change: We consider seven speed changes ranging from .25x to 2x times the speed of an original clip with an increment of .25x, skipping 1x since that would represent the original clip, which we have already included as a pitch shift with value $0$.
    \item Noise Superposition: We consider 10 randomly picked vacuum  and washing machine sound clips, obtained from \rule{0pt}{7ex} the environmental sound classification dataset~\cite{ESC50}, as background noises to modify original breathing event clips with eight different signal-to-noise ratio levels ranging from \ignore{0.0001}$10^{-4}$ to \ignore{10000}$10^{4}$, incremented by magnitudes of 10 while skipping $1$.
\end{itemize}
Thereby, each original breathing clip is modified 102 times.

\subsection{Feature Computation}\label{f_comp}

We compute the following sets of candidate features.
\begin{itemize}
    \item Heart rate features: From the windows of 10 samples we compute 21 statistical features\ignore{: mean, median, standard deviation, variance, coefficient of variance, range, coefficient of range, first quartile or $25^{th}$ percent, third quartile or $75^{th}$ percent, max, interquartile range, coefficient of interquartile, mean absolute deviation, median absolute deviation, energy, power, root mean square, root sum of squares, signal to noise ratio, skewness, and kurtosis,} described in our previous work~\cite{vhaduri2019multi}.
    \item Gait features: We compute the same above mentioned 21 features from each window of x-, y-, and z-axis readings obtained from both gyroscope and accelerometer. 
    \item Audio features: From each inhalation breathing event (original and augmented), we compute 40 Mel-frequency cepstral coefficients (MFCCs)\ignore{, where, $MFCCi$ represents the $i^{th}$ coefficient}.
\end{itemize}
Thereby, we obtain 21 and 126 (21 from each of the six axes) features from a single window of heart rate and gait data, respectively, and 40 features from every breathing clip.

\subsection{Feature Selection}\label{f_select}

To select the most influential features, we use the Sci-kit learn feature selection package ``Select the $K$ Best Features'' (SelectKBest), which provides an importance score for each feature and based on that score we rank the features. We try with different numbers of features, i.e., $K$, to find the best model performance. 
In this work, we find $K = 20$ performs the best. 
In each iteration of the leave-one-out validation\ignore{training-testing}, described in Section~\ref{train-test}, we select different feature sets, which are very similar with changes in ordering.

\begin{table*}[htbp]\rule{0pt}{4ex}  
\caption{The best HR models with average and standard deviation of performance measures}
\label{H_Results}
\centering
\begin{tabular}{l|c|c|c|c|c|c|c} 
\hline
BINARY Model    \\
\hline
Classifier (parameters) & feature count & ACC & RMSE & FAR & FRR & $F_1$ score & AUC-ROC \\
\hline
RF (n estimators = 450) & 20 & 0.64 (0.12) & 0.04 (0.01) & 0.30 (0.15) & 0.42 (0.16) & 0.61 (0.14) & 0.64 (0.12)\\
\hline
$k$-NN ($k=32$, minkowski distance) & 20 & 0.63 (0.11) & 0.04 (0.01) & 0.37 (0.15) & 0.36 (0.14) & 0.63 (0.12) & 0.63 (0.11)\\
\hline
NB & 20 & 0.65 (0.11) & 0.04 (0.01) & 0.36 (0.25) & 0.39 (0.19) & 0.61 (0.12) & 0.63 (0.11)\\
\hline
\rowcolor{lightgray} SVM (RBF kernel,$\gamma=0.03$, $C=3$) & 20 & 0.66 (0.11) & 0.04 (0.01) & 0.29 (0.16) & 0.38 (0.17) & 0.63 (0.14) & 0.66 (0.11)\\
\hline
SVM (Poly. kernel, $d=1$, $C=1$) & 20 & 0.65 (0.12) & 0.04 (0.01) & 0.26 (0.20) & 0.44 (0.23) & 0.59 (0.18) & 0.65 (0.12)\\
\hline 
UNARY Model    \\
\hline
SVM (RBF kernel, $\gamma=0.05$, $\nu=0.5$) & 20 & 0.56 (0.08) &  0.05 (0.00) & 0.41 (0.14) & 0.46 (0.09) & 0.55 (0.08) & N/A \\
\hline
\end{tabular}
\end{table*}

\begin{table*}[!t]
\caption{The best HRG models with average and standard deviation of performance measures}
\label{HG_Results}
\centering
\begin{tabular}{l|c|c|c|c|c|c|c} 
\hline
BINARY Model    \\
\hline
Classifier (parameters) & feature count & ACC & RMSE & FAR & FRR & $F_1$ score & AUC-ROC \\
\hline
RF (n estimators = 450) & 20 & 0.69 (0.13) & 0.04 (0.01) & 0.47 (0.32) & 0.15 (0.21) & 0.73 (0.21) & 0.71 (0.13)\\
\hline
$k$-NN ($k=24$, minkowski distance) & 20 & 0.79 (0.07) & 0.03 (0.01) & 0.19 (0.10) & 0.23 (0.09) & 0.79 (0.08) & 0.79 (0.07)\\
\hline
NB & 20 & 0.65 (0.10) & 0.04 (0.01) & 0.28 (0.26) & 0.42 (0.27) & 0.62 (0.20) & 0.66 (0.10)\\
\hline
\rowcolor{lightgray} SVM (RBF kernel,$\gamma=0.05$, $C=5$) & 20 & 0.82 (0.08) & 0.03 (0.01) & 0.17 (0.09) & 0.19 (0.10) & 0.81 (0.08) & 0.82 (0.08)\\
\hline
SVM (Poly. kernel, $d=3$, $C=14$) & 20 & 0.78 (0.09) & 0.03 (0.01) & 0.19 (0.12) & 0.25 (0.13) & 0.77 (0.10) & 0.78 (0.09)\\
\hline 
UNARY Model    \\
\hline
SVM (RBF kernel, $\gamma=0.05$, $\nu=0.5$) & 20 & 0.72 (0.10) &  0.04 (0.01) & 0.28 (0.16) & 0.29 (0.08) & 0.72 (0.09) & N/A \\
\hline
\end{tabular}
\end{table*}

\begin{table*}[!t]
\caption{The best HRB models with average and standard deviation of performance measures}
\label{HS_Results}
\centering
\begin{tabular}{l|c|c|c|c|c|c|c} 
\hline
BINARY Model    \\
\hline
Classifier (parameters) & feature count & ACC & RMSE & FAR & FRR & $F_1$ score & AUC-ROC \\
\hline
RF (n estimators = 600) & 20 & 0.90 (0.07) & 0.02 (0.01) & 0.13 (0.10) & 0.07 (0.08) & 0.90 (0.07) & 0.90 (0.07)\\
\hline
$k$-NN ($k=2$, minkowski distance) & 20 & 0.92 (0.07) & 0.02 (0.01) & 0.08 (0.07) & 0.09 (0.11) & 0.91 (0.09) & 0.92 (0.07)\\
\hline
NB & 20 & 0.75 (0.05) & 0.04 (0.00) & 0.22 (0.10) & 0.29 (0.12) & 0.73 (0.07) & 0.75 (0.05)\\
\hline
\rowcolor{lightgray} SVM (RBF kernel,$\gamma=0.08$, $C=4$) & 20 & 0.94 (0.07) & 0.02 (0.01) & 0.06 (0.07) & 0.07 (0.09) & 0.93 (0.08) & 0.94 (0.07)\\
\hline
SVM (Poly. kernel, $d=4$, $C=16$) & 20 & 0.91 (0.07) & 0.02 (0.01) & 0.06 (0.06) & 0.11 (0.08) & 0.91 (0.07) & 0.91 (0.07)\\
\hline 
UNARY Model    \\
\hline
SVM (RBF kernel, $\gamma=0.05$, $\nu=0.5$) & 20 & 0.72 (0.07) &  0.04 (0.00) & 0.32 (0.10) & 0.24 (0.06) & 0.73 (0.06) & N/A \\
\hline
\end{tabular}
\end{table*}

\subsection{Methods}\label{methods}

\begin{figure}[!b]
    \centering
    \scalebox{.17}{\includegraphics{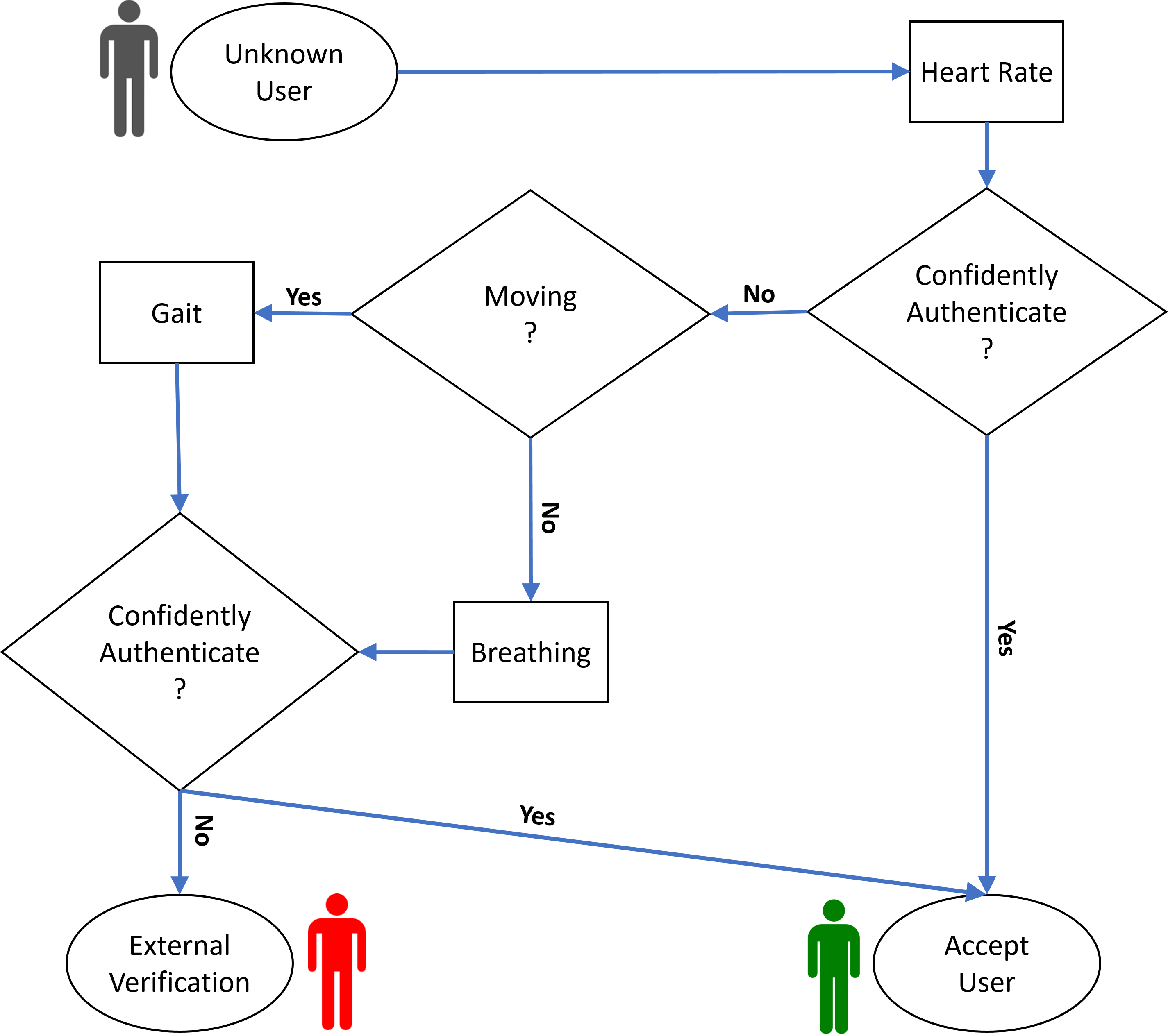}}
    \caption{\textcolor{black}{Proposed wearable device user authentication scheme}}
    \label{Bio Flow}
\end{figure}

In Figure \ref{Bio Flow}, we present an overview of our proposed implicit and continuous wearable-user authentication scheme using person-dependent multiple biometrics that are readily available on most of the wearables in the market. Depending on a user's context, i.e., user's state, various routes of the authentication scheme will be executed.  

We first try to authenticate a user based on the heart rate obtained from the photo-plethysmogram (PPG) sensors since this biometric data is always available irrespective of a user's state. However, the heart rate data may not be precised to identify the user when it is recorded in coarse-grained, e.g., one samples per minute. Additionally, the heart rate biometric can be easily affected by different factors, such as motion artifacts or stress. Therefore, if the system cannot authenticate the user with enough confidence, it checks the next authentication module that relies on other biometrics. \ignore{Otherwise, the system allows the user to access the device.}

The authentication system first tries to check whether the user is moving utilizing the on-device accelerometer and gyroscope data. If the user is moving, the system tries to authenticate the user based on gait and heart rate biometrics \rule{0pt}{7ex}  together. If the system can authenticate the user with enough confidence, it allows the user to access the device. 

However, if the user does not move or the gait and heart rate-based module cannot authenticate the user, the system tries to combine breathing biometrics collected from the on-device microphone. 
During sedentary states, audio recordings from wearables are less affected by motion artifacts. Thereby, the breathing audio recordings could be a good biometric to identify users, while during sedentary states. 
If the system can authenticate the user with enough confidence, it allows the user to access the device. 
Otherwise, the user's access to the device is revoked and require some sort of external verification, such as pin locks or passwords.


Based on the combination of the three biometrics that we use in our authentication approach, we define the following models:
\begin{itemize}
\item
Heart rate data-driven model (HR model)
\item
Heart rate and gait data-driven model (HRG model)
\item
Heart rate and breathing data-driven model (HRB model)
\ignore{***
\item
Heart rate, gait, and breathing data-driven model (HRGB model)
***}
\end{itemize}

While developing the above models, we consider various classifiers, including the $k$-nearest neighbor ($k$-NN), random forest (RF), naive bayes (NB), and support vector machine (SVM) with binary and unary schemes\ignore{ using the Sci-kit learn libraries}. Compared to binary, unary models are available only for the SVM classifiers with radial basis function (RBF) kernels. 

Based on the windowing approach discussed in Section~\ref{vData}, we can derive the time complexity of our authentication system based on the sampling frequency of different sensors. For example, let us consider a case where every heart rate sample is collected in $x$ seconds. With this sampling frequency, it will take $10x$ seconds to make a window of 10 heart rate samples. Therefore, if the HR model can successfully validate a user, it will take $10x$ seconds to complete the authentication process. However, if the HR models to validate a user, the system can take one of the two paths based on the user's context. If the user is in a {\em non-sedentary} state, then the HRG model will be triggered, which will wait for an additional $10x$ seconds to collect 10 gait and heart rate samples; thereby, a total of $20x$ seconds will be required for the system to validate the user. 
If the HR model fails to validate the user and the user's context, i.e., physical state is {\em sedentary}, then the system will try to authenticate the user based on breathing events in addition to heart rate. Since the average length of a breathing events used in this work is 1.4 seconds; therefore, if $x >= 0.14$ seconds (i.e., a single heart rate window is longer than the breathing event), the system will need $10x$ seconds to gather 10 new heart rate samples for the HRB model to test the user. Thereby, it will take in total $20x$ seconds for the system to authenticate the user. 
However, if $x < 0.14$ seconds, i.e., breathing events are larger than the heart rate windows, the system will take $10x + 1.4$ seconds to validate the user.

\section{User Authentication}
\label{authentication}

Before presenting the detailed evaluation of our models, we first present training-testing set split and our modeling schemes, followed by list of performance measures and hyper-parameter optimization.  

\subsection{Training-Testing Set}\label{train-test}

In our binary modeling, we try to distinguish a valid user (class-$0$) from the impostors (class-$1$). To avoid overfitting, we consider at least 10 times more feature windows, i.e., number of instances than the number of features. While training-testing, we follow the leave-one-out strategy, where we train-test $N$ unique models one-by-one for each user with $N$ number of instances. During each training-testing, we keep one instance for testing and use the rest of the $N-1$ instances for training\ignore{ with $N$ be the number of total instances from each class}. Since we have 10 subjects and perform 10 leave-one-out testing for each subject; thereby, all aggregated performance measures presented in this paper are based on 60 performance measures.

For class balancing, in case of binary models, we consider the same $N-1$ number of instances from each class. Since our imposter class (class-$1$) consists of nine person data (i.e., all subjects except the one considered as valid subject or class-$0$), we pick $(N-1)/9$ instances from each imposter. 
For example, while training a HR model, we consider 510 heart rate windows from a target/valid user and 510/9 $\approx$ 56 windows from each of the nine imposters. In the test set, we consider 102 windows from the valid user and 102/9 $\approx$ 11 windows from each imposter. Similarly, while training a HRB model, we use 510 windows, i.e., breathing events from a valid user in addition to 510 heart rate windows. \rule{0pt}{7ex} Where, 510 breathing events are obtained from the five original breathing events and their 102 augmented events, i.e., $5 \times 102$ = 510. To keep the training and test set separate, to use the remaining one breathing event and its 102 augmented events, i.e., 102 events/windows. For imposter, we uniformly select the windows to ensure a balanced classification.
In case of unary models, we also follow the leave-one-out strategy. But, compared to the binary, unary models are developed with only a valid user's data with an outlier rate ($\nu$) to split the user's data into valid and outlier groups. In case of our experiments, we find $\nu = 0.5$ as the optional outlier rate. 

\begin{figure}[!b]
\centering
\begin{subfigure}{.45\textwidth}
  \centering
  \includegraphics[width=.95\linewidth]{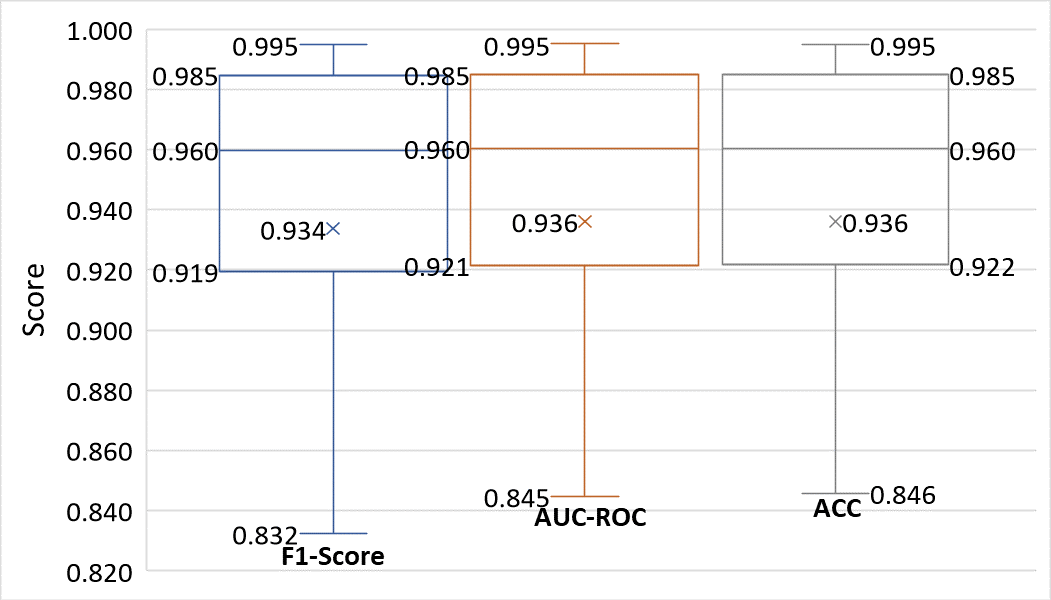}
  \caption{}
  \label{pos_binary}
\end{subfigure}%
\\
\begin{subfigure}{.45\textwidth}
  \centering
  \includegraphics[width=.95\linewidth]{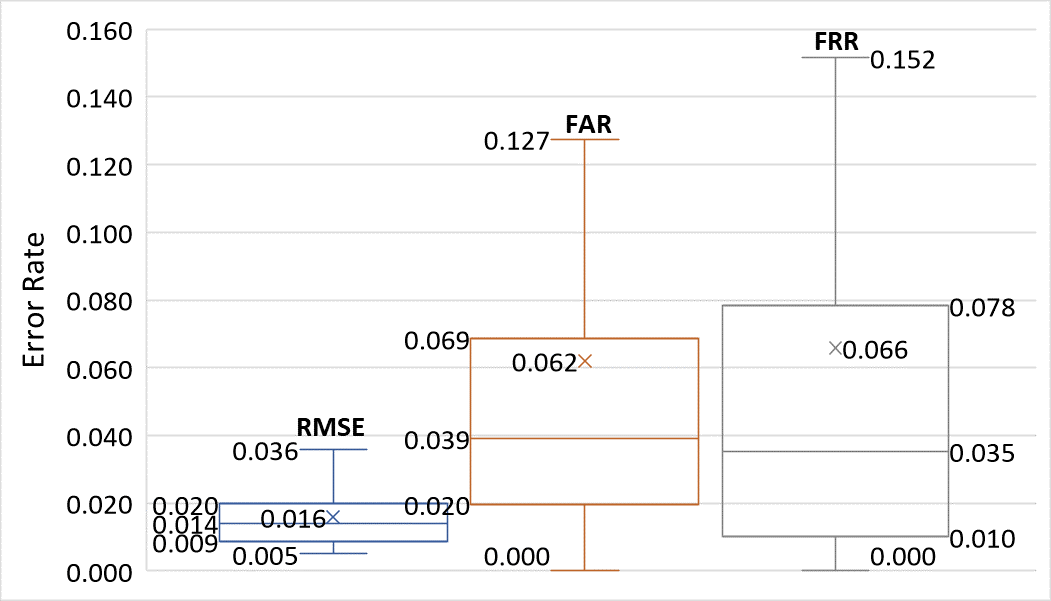}
  \caption{}
  \label{neg_binary}
\end{subfigure}%
\caption{Box plots of (a) positive and (b) negative measures of performance of the HRB model with Binary SVM RBF classifier. Cross markers ($\times$) represent the average values.}
\label{boxes_binary}
\end{figure}

\subsection{Performance Measures}\label{perMeasures}

\ignore{***************************************
To evaluate the performance of different models we consider 
Accuracy (ACC), 
Root Mean Square Error (RMSE),
False Acceptance Rate (FAR), 
False Rejection Rate (FRR),
$F_1$ score, Area Under the Curve - Receiver Operating Characteristic (AUC-ROC), and Equal Error Rate (EER).
Terminologies have their usual meaning in machine learning, when classifying a subject using a feature set~\cite{vhaduri2019multi,vhaduri2014estimating}. 
For an ideal system, it is desirable to have a lower RMSE, EER, FAR and FRR, but a higher ACC, $F_1$ score, and AUC-ROC. 
***************************************}

To evaluate the performance of different modeling approaches, we consider the following measures:\\

{\bf {\em Accuracy (ACC)}}, which is the fraction of predictions that are correct, i.e., 
\begin{equation}
\label{acc}
ACC = \frac{TP+TN}{TP+FN+FP+TN} 
\end{equation}

{\bf {\em Root Mean Square Error (RMSE)}}, which is the square root of the sum of squares of the deviation from the prediction to \rule{0pt}{7ex}the actual value. It is equivalent to the square root of the rate of misclassification, i.e., 
\begin{equation}
\label{rmse}
RMSE = \sqrt{\frac{FP+FN}{TP+FN+FP+TN}}
\end{equation}

{\bf {\em  False Acceptance Rate (FAR)}}, which is the fraction of invalid users accepted by an authentication system, i.e.:
\begin{equation}
\label{FAR}
FAR = \frac{FP}{FP+TN} 
\end{equation}

{\bf {\em  False Rejection Rate (FRR)}}, which is the fraction of genuine users rejected by an authentication system, i.e.:
\begin{equation}
\label{FRR}
FRR = \frac{FN}{TP+FN} 
\end{equation}

{\bf {\em  $F_1$ Score}}, which is the measure of performance of an authentication system based on both it precision (positive predictive value) and recall (true positive rate) measures, i.e.:
\begin{equation}
\label{f1score}
F_1 Score = 2\left(\frac{TP}{TP+FN}+\frac{TP}{TP+FP}\right)^{-1}
\end{equation}

{\bf {\em  Area Under the Curve - Receiver Operating Characteristic (AUC-ROC)}}, which is the graphical relationship between FAR and FRR with the change of thresholds.
Where terminologies used in Equations~\ref{acc},~\ref{rmse},~\ref{FAR},~\ref{FRR}, and~\ref{f1score} have their usual meaning in machine learning, when classifying a subject using a feature set.  
Therefore, a desirable authentication system should have lower negative measures (i.e., RMSE, FAR, and FRR), but higher positive measures (i.e., ACC, $F_1$ Score, and AUC-ROC) of performance. 
We also use {\bf {\em Equal Error Rate} (EER)}, which is defined as the point when FRR and FAR are equal, i.e., a trade-off between the two error measures (i.e., FRR and FAR)\ignore{ \textcolor{green}{and with the $k$-NN Heart Rate model shown in figure \ref{Error Rate HR} the EER is at the confidence threshold of 49\%}.} 

\ignore{---------------------------------------------
\begin{figure}[H]
\centering
\begin{subfigure}{.47\textwidth}
  \centering
  \includegraphics[width=.95\linewidth]{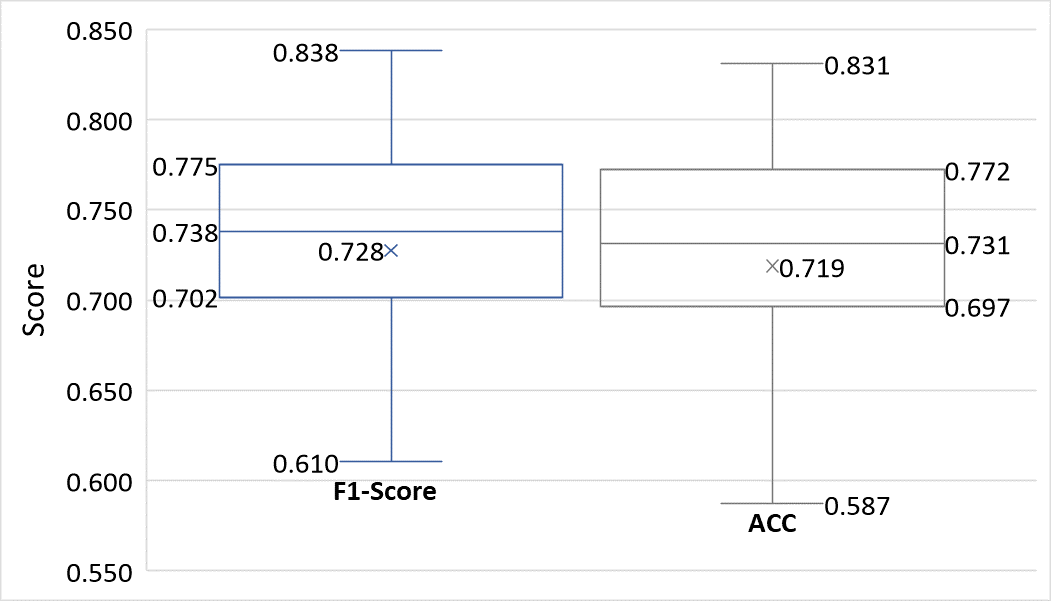}
  \caption{}
  \label{pos_unary}
\end{subfigure}%
\\
\begin{subfigure}{.47\textwidth}
  \centering
  \includegraphics[width=.95\linewidth]{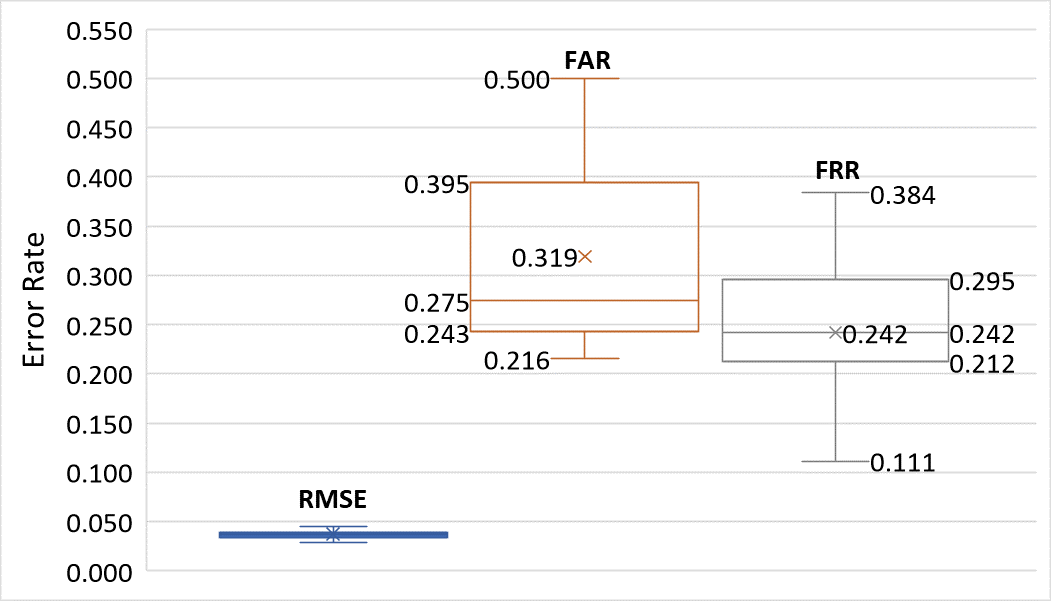}
  \caption{}
  \label{neg_unary}
\end{subfigure}%
\caption{Box plots of (a) positive and (b) negative measures of performance of the HRB model with Unary SVM RBF classifier. Cross markers ($\times$) represent the average values.}
\label{boxes_unary}
\end{figure}
---------------------------------------------}

\subsection{Hyper-Parameter Optimization}\label{param_opt}

We use the grid search package in the Sci-kit learn to find the optimal hyper-parameter sets. For each leave-one-out modeling, we separately perform the hyper-parameter optimization using various ranges of values.
From the different iterations of the leave-one-out approach, we obtain similar values for the hyper-parameters.  
\ignore{the displayed optimal hyper-parameters are those of the best ($F_1$ score) iteration for simplicity.}
In Tables~\ref{H_Results},~\ref{HG_Results}, and~\ref{HS_Results}, we present the set of optimal values obtained from different modeling approaches.

\subsection{Authentication Model Evaluation}\label{model_performance}

In Tables~\ref{H_Results},~\ref{HG_Results}, and~\ref{HS_Results}, we present the performance of the best models using various biometric combinations and different classifiers. 
In Table \ref{H_Results}, we observe that the best binary HR model (i.e., model that only uses heart rate data) can provide an average ACC and AUC-ROC of $0.66 \pm 0.11$. 
As discussed previously in the Section~\ref{methods}, if the HR model is not confident enough to authenticate a user or fails to authenticate, we use additional biometrics, such as, gait or breathing sound. 
Compared to binary, for the unary HR model, we observe low performance, i.e., an average ACC of $0.56 \pm 0.08$\ignore{ suffers a lot from the lack of imposter training data with only} since Unary model considers portions of a valid user's data as outliers.

\begin{figure}[!b]
    \centering
    \scalebox{.35}{\includegraphics{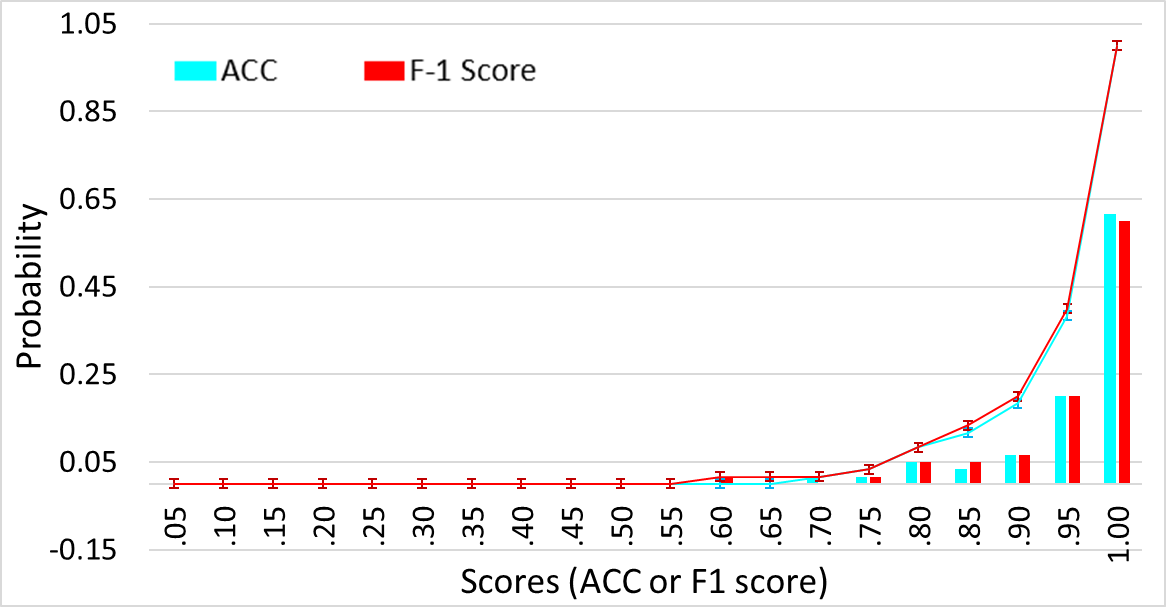}}
    \caption{PDF and CDF with error bars of binary HRB SVM (RBF) model performance.}
    \label{pdf_cdf_binary}
\end{figure} 

\begin{figure}[!b]
    \centering
    \scalebox{.5}{\includegraphics{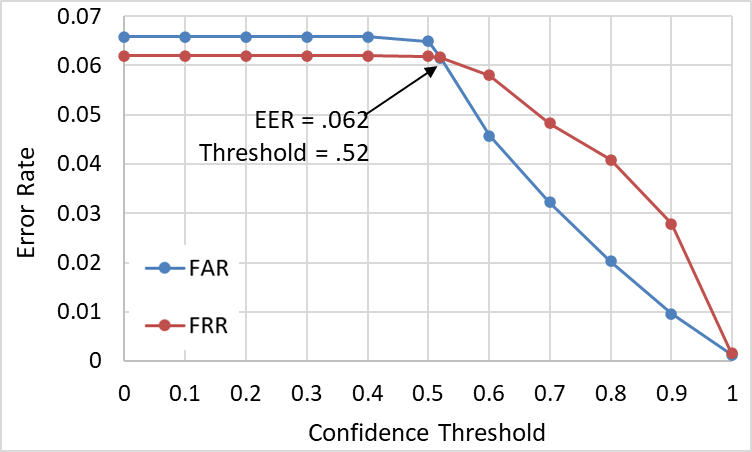}}
    \caption{Change of error rates with varying confidence thresholds using the binary HRB SVM (RBF) model.}
    \label{Binary Confidence}
\end{figure}

\rule{0pt}{4ex} In Table~\ref{HG_Results}, we observe that adding gait biometric (when available) with heart rate, all measures improve. In case of the best binary HRG model (i.e., model that uses heart rate and gait biometrics), ACC and AUC-ROC increased by 24\%; $F_1$ score increased by 29\% compared to the best binary HR model. 
The FAR also improves (i.e., drops) from $0.29 \pm 0.16$ to $0.17\pm0.09$. Though gait data is only available while a user is moving, its addition to less accurate minute-level heart rate data can significantly improve authentication performance. Similarly to binary, the unary HRG model shows promise over the unary HR model with an overall increase of about 29\% both for ACC and $F_1$ score. 

In Table~\ref{HS_Results}, we observe that the HRB model (i.e., model that uses heart rate and breathing biometrics) achieves a better performance compared to the HRG model. We achieve a 65\% drop in the FAR while comparing the binary HRB  with the binary HRG model. Additionally, we observe $\approx$ 15\% increase, while comparing the ACC, $F_1$ score, and AUC-ROC of the binary HRB model with the binary HRG model. 
While comparing the HRB model to the HR model, we observe a huge performance improvement. Compared to the binary HR model, the binary HRB model performs better in terms of $F_1$ score (an increase of 48\%) and AUC-ROC (an increase of 42\%) with a high accuracy of $0.94 \pm 0.07$. The unary HRB model performs similar to the unary HRG model with a lower standard deviation, i.e., higher consistency, in terms of ACC ($0.10$ vs. $0.07$) and $F_1$ score ($0.09$ vs. $0.06$).

\ignore{-----------------------
\begin{figure}[ht]
    \centering
    \scalebox{.425}{\includegraphics{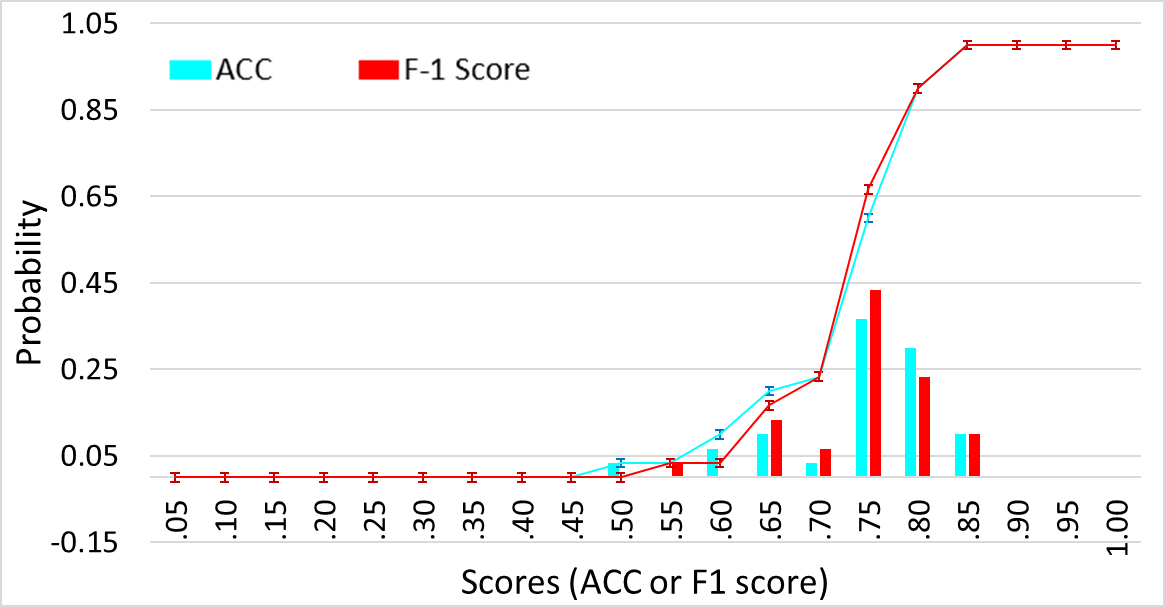}}
    \caption{PDF and CDF with error bars of unary HRB SVM (RBF) model performance}
    \label{pdf_cdf_unary}
\end{figure}
------------------------------}

\ignore{
In Figure~\ref{boxes_binary}\ignore{ and~\ref{boxes_unary}}, we present the summary of different performance measures. In addition to the five measures of a box plot, we also present the average value.}

In Figure~\ref{boxes_binary}\ignore{ and~\ref{boxes_unary}}, we present five summarized values of different performance measures in addition to the average values \rule{0pt}{5ex} presented in Table~\ref{HS_Results}. In the figure, we observe that median of each performance measure is better than average, since average is easily affected by outliers, which we do not show for the simplicity of visualization. 
For example, we obtain 2.6\% better ACC, 2.8\% better $F_1$ score, 59\% better FAR, and 89\% better FRR, while comparing median with average values.
\ignore{In Figure~\ref{pos_binary}}Additionally, we observe that the interquartile ranges of\ignore{ ACC, $F_1$ score, and AUC-ROC} different performance measures are about 0.07 (Figure~\ref{pos_binary}) and 0.05 (Figure~\ref{neg_binary}). 
Similarly, \ignore{unary models in Figures~\ref{pos_unary} and~\ref{neg_unary} show}in the case of unary modeling, we obtain tighter interquartile ranges\ignore{ and that even though Figure \ref{neg_unary} shows FAR on the higher end we can mitigate that with a higher confidence threshold}.
These narrow interquartile ranges represent the consistency of performance measures\ignore{ regardless of valid user}. 

In Figure~\ref{pdf_cdf_binary}\ignore{ and~\ref{pdf_cdf_unary}}, we present the Probability Distribution Function (PDF) and Cumulative Distribution Function (CDF) with error bars of performance of the best binary\ignore{ and unary models, respectively} model.
In Figure \ref{pdf_cdf_binary}, around 65\% of the performance values (both ACC and $F_1$ scores) fall in the range of 0.95--1, which shows that our models perform very well for the most of the cases. 
\ignore{In Figure \ref{pdf_cdf_unary}}In the case of unary modeling, we observe that \ignore{majority ($\approx$ 66\%)} $\frac{2}{3}$ (i.e., $\approx$ 66\%) of the values fall in the range of 0.7--0.8, which is also a reasonable performance for unary model~\cite{vhaduri2019multi}. 

Additionally, in the figure, we observe that\ignore{ both binary and unary} from the errors bars are very short, i.e., achieved performance values are highly consistent. Therefore, our developed models consistently perform well.

\subsection{Error Analysis}\label{accessing_error_rate}
In this section, we present an analysis on how our system performs with the change of confidence levels, i.e., thresholds. In case of an ideal system, it is desired to have a lower FAR and FRR. 
In Figures~\ref{Binary Confidence}, we present our analysis of error rates (FAR and FRR) with varying confidence thresholds for the binary HRB SVM (RBF) model. We observe that at confidence threshold 0.52 FAR and FRR intersects with an equal error rate (EER) of about 0.06. 
After this point, error rates drop quickly. We observe that FAR and FRR drops below 0.05 after threshold values around 0.6 and 0.7, respectively. 

\section{Limitations, Discussion, and Conclusions}

To the best of our knowledge, this is the first work that attempts to authenticate a wearable device user without any explicit user\ignore{ burden} interaction utilizing three easily obtainable soft-biometrics (i.e., heart rate, gait, and breathing sounds) in a more context-based approach, i.e., availability of data. 
\ignore{Through our detailed analysis we show that w}We can authenticate a user with an average accuracy and AUC-ROC of $0.94\pm0.07$, $F_1$ score of $0.93\pm0.08$, and an EER of about 0.06 FAR at 0.52 confidence threshold while considering the heart rate and breathing sounds.
This shows the promise to develop a continuous implicit-authentication system for the market wearables utilizing their limited sensing and computational capabilities\ignore{ in order to secure our valuable information as well as to create a safe gateway to unlock cars, access online accounts, etc}.  

This work has some limitations, which we plan to address in the future. First, we have limited number of audio breathing clips. However, we increase the data volume using standard \rule{0pt}{6ex} audio augmentation approaches. Second, in this feasibility work, we use a set of ten subjects. However, we perform a leave-one-out validation approach\ignore{ to deal with this limitation} and our achieved performance shows a promise to further investigate this with a large-scale extended period study. Third, we use different datasets, which could affect the performance. However, we use three independent biometrics and perform feature selection analysis to optimize implementation; thereby, our results potentially shows a baseline performance, which could further be improved by using the three biometrics from the same subject since that could more robustly identify a user compared to our case.  
Finally, more advanced modeling techniques such as deep learning models (recurrent neural networks or convolutional neural networks) may further improve the accuracy of the models, but that will require to off load data from the wearable\ignore{ to server}, which can lead to additional security challenges; therefore, our approach \ignore{of using traditional machine learning models have}has a higher scope to implement on the wearables.

\bibliographystyle{IEEEtran}
\bibliography{reference}

\ignore{----------------------------------------------------
\section{Introduction}
This document is a model and instructions for \LaTeX.
Please observe the conference page limits. 

\section{Ease of Use}

\subsection{Maintaining the Integrity of the Specifications}

The IEEEtran class file is used to format your paper and style the text. All margins, 
column widths, line spaces, and text fonts are prescribed; please do not 
alter them. You may note peculiarities. For example, the head margin
measures proportionately more than is customary. This measurement 
and others are deliberate, using specifications that anticipate your paper 
as one part of the entire proceedings, and not as an independent document. 
Please do not revise any of the current designations.

\section{Prepare Your Paper Before Styling}
Before you begin to format your paper, first write and save the content as a 
separate text file. Complete all content and organizational editing before 
formatting. Please note sections \ref{AA}--\ref{SCM} below for more information on 
proofreading, spelling and grammar.

Keep your text and graphic files separate until after the text has been 
formatted and styled. Do not number text heads---{\LaTeX} will do that 
for you.

\subsection{Abbreviations and Acronyms}\label{AA}
Define abbreviations and acronyms the first time they are used in the text, 
even after they have been defined in the abstract. Abbreviations such as 
IEEE, SI, MKS, CGS, ac, dc, and rms do not have to be defined. Do not use 
abbreviations in the title or heads unless they are unavoidable.

\subsection{Units}
\begin{itemize}
\item Use either SI (MKS) or CGS as primary units. (SI units are encouraged.) English units may be used as secondary units (in parentheses). An exception would be the use of English units as identifiers in trade, such as ``3.5-inch disk drive''.
\item Avoid combining SI and CGS units, such as current in amperes and magnetic field in oersteds. This often leads to confusion because equations do not balance dimensionally. If you must use mixed units, clearly state the units for each quantity that you use in an equation.
\item Do not mix complete spellings and abbreviations of units: ``Wb/m\textsuperscript{2}'' or ``webers per square meter'', not ``webers/m\textsuperscript{2}''. Spell out units when they appear in text: ``. . . a few henries'', not ``. . . a few H''.
\item Use a zero before decimal points: ``0.25'', not ``.25''. Use ``cm\textsuperscript{3}'', not ``cc''.)
\end{itemize}

\subsection{Equations}
Number equations consecutively. To make your 
equations more compact, you may use the solidus (~/~), the exp function, or 
appropriate exponents. Italicize Roman symbols for quantities and variables, 
but not Greek symbols. Use a long dash rather than a hyphen for a minus 
sign. Punctuate equations with commas or periods when they are part of a 
sentence, as in:
\begin{equation}
a+b=\gamma\label{eq}
\end{equation}

Be sure that the 
symbols in your equation have been defined before or immediately following 
the equation. Use ``\eqref{eq}'', not ``Eq.~\eqref{eq}'' or ``equation \eqref{eq}'', except at 
the beginning of a sentence: ``Equation \eqref{eq} is . . .''

\subsection{\LaTeX-Specific Advice}

Please use ``soft'' (e.g., \verb|\eqref{Eq}|) cross references instead
of ``hard'' references (e.g., \verb|(1)|). That will make it possible
to combine sections, add equations, or change the order of figures or
citations without having to go through the file line by line.

Please don't use the \verb|{eqnarray}| equation environment. Use
\verb|{align}| or \verb|{IEEEeqnarray}| instead. The \verb|{eqnarray}|
environment leaves unsightly spaces around relation symbols.

Please note that the \verb|{subequations}| environment in {\LaTeX}
will increment the main equation counter even when there are no
equation numbers displayed. If you forget that, you might write an
article in which the equation numbers skip from (17) to (20), causing
the copy editors to wonder if you've discovered a new method of
counting.

{\BibTeX} does not work by magic. It doesn't get the bibliographic
data from thin air but from .bib files. If you use {\BibTeX} to produce a
bibliography you must send the .bib files. 

{\LaTeX} can't read your mind. If you assign the same label to a
subsubsection and a table, you might find that Table I has been cross
referenced as Table IV-B3. 

{\LaTeX} does not have precognitive abilities. If you put a
\verb|\label| command before the command that updates the counter it's
supposed to be using, the label will pick up the last counter to be
cross referenced instead. In particular, a \verb|\label| command
should not go before the caption of a figure or a table.

Do not use \verb|\nonumber| inside the \verb|{array}| environment. It
will not stop equation numbers inside \verb|{array}| (there won't be
any anyway) and it might stop a wanted equation number in the
surrounding equation.

\subsection{Some Common Mistakes}\label{SCM}
\begin{itemize}
\item The word ``data'' is plural, not singular.
\item The subscript for the permeability of vacuum $\mu_{0}$, and other common scientific constants, is zero with subscript formatting, not a lowercase letter ``o''.
\item In American English, commas, semicolons, periods, question and exclamation marks are located within quotation marks only when a complete thought or name is cited, such as a title or full quotation. When quotation marks are used, instead of a bold or italic typeface, to highlight a word or phrase, punctuation should appear outside of the quotation marks. A parenthetical phrase or statement at the end of a sentence is punctuated outside of the closing parenthesis (like this). (A parenthetical sentence is punctuated within the parentheses.)
\item A graph within a graph is an ``inset'', not an ``insert''. The word alternatively is preferred to the word ``alternately'' (unless you really mean something that alternates).
\item Do not use the word ``essentially'' to mean ``approximately'' or ``effectively''.
\item In your paper title, if the words ``that uses'' can accurately replace the word ``using'', capitalize the ``u''; if not, keep using lower-cased.
\item Be aware of the different meanings of the homophones ``affect'' and ``effect'', ``complement'' and ``compliment'', ``discreet'' and ``discrete'', ``principal'' and ``principle''.
\item Do not confuse ``imply'' and ``infer''.
\item The prefix ``non'' is not a word; it should be joined to the word it modifies, usually without a hyphen.
\item There is no period after the ``et'' in the Latin abbreviation ``et al.''.
\item The abbreviation ``i.e.'' means ``that is'', and the abbreviation ``e.g.'' means ``for example''.
\end{itemize}
An excellent style manual for science writers is \cite{b7}.

\subsection{Authors and Affiliations}
\textbf{The class file is designed for, but not limited to, six authors.} A 
minimum of one author is required for all conference articles. Author names 
should be listed starting from left to right and then moving down to the 
next line. This is the author sequence that will be used in future citations 
and by indexing services. Names should not be listed in columns nor group by 
affiliation. Please keep your affiliations as succinct as possible (for 
example, do not differentiate among departments of the same organization).

\subsection{Identify the Headings}
Headings, or heads, are organizational devices that guide the reader through 
your paper. There are two types: component heads and text heads.

Component heads identify the different components of your paper and are not 
topically subordinate to each other. Examples include Acknowledgments and 
References and, for these, the correct style to use is ``Heading 5''. Use 
``figure caption'' for your Figure captions, and ``table head'' for your 
table title. Run-in heads, such as ``Abstract'', will require you to apply a 
style (in this case, italic) in addition to the style provided by the drop 
down menu to differentiate the head from the text.

Text heads organize the topics on a relational, hierarchical basis. For 
example, the paper title is the primary text head because all subsequent 
material relates and elaborates on this one topic. If there are two or more 
sub-topics, the next level head (uppercase Roman numerals) should be used 
and, conversely, if there are not at least two sub-topics, then no subheads 
should be introduced.

\subsection{Figures and Tables}
\paragraph{Positioning Figures and Tables} Place figures and tables at the top and 
bottom of columns. Avoid placing them in the middle of columns. Large 
figures and tables may span across both columns. Figure captions should be 
below the figures; table heads should appear above the tables. Insert 
figures and tables after they are cited in the text. Use the abbreviation 
``Fig.~\ref{fig}'', even at the beginning of a sentence.

\begin{table}[htbp]
\caption{Table Type Styles}
\begin{center}
\begin{tabular}{|c|c|c|c|}
\hline
\textbf{Table}&\multicolumn{3}{|c|}{\textbf{Table Column Head}} \\
\cline{2-4} 
\textbf{Head} & \textbf{\textit{Table column subhead}}& \textbf{\textit{Subhead}}& \textbf{\textit{Subhead}} \\
\hline
copy& More table copy$^{\mathrm{a}}$& &  \\
\hline
\multicolumn{4}{l}{$^{\mathrm{a}}$Sample of a Table footnote.}
\end{tabular}
\label{tab1}
\end{center}
\end{table}

\begin{figure}[htbp]
\centerline{\includegraphics{fig1.png}}
\caption{Example of a figure caption.}
\label{fig}
\end{figure}

Figure Labels: Use 8 point Times New Roman for Figure labels. Use words 
rather than symbols or abbreviations when writing Figure axis labels to 
avoid confusing the reader. As an example, write the quantity 
``Magnetization'', or ``Magnetization, M'', not just ``M''. If including 
units in the label, present them within parentheses. Do not label axes only 
with units. In the example, write ``Magnetization (A/m)'' or ``Magnetization 
\{A[m(1)]\}'', not just ``A/m''. Do not label axes with a ratio of 
quantities and units. For example, write ``Temperature (K)'', not 
``Temperature/K''.

\section*{Acknowledgment}

The preferred spelling of the word ``acknowledgment'' in America is without 
an ``e'' after the ``g''. Avoid the stilted expression ``one of us (R. B. 
G.) thanks $\ldots$''. Instead, try ``R. B. G. thanks$\ldots$''. Put sponsor 
acknowledgments in the unnumbered footnote on the first page.

\section*{References}

Please number citations consecutively within brackets \cite{b1}. The 
sentence punctuation follows the bracket \cite{b2}. Refer simply to the reference 
number, as in \cite{b3}---do not use ``Ref. \cite{b3}'' or ``reference \cite{b3}'' except at 
the beginning of a sentence: ``Reference \cite{b3} was the first $\ldots$''

Number footnotes separately in superscripts. Place the actual footnote at 
the bottom of the column in which it was cited. Do not put footnotes in the 
abstract or reference list. Use letters for table footnotes.

Unless there are six authors or more give all authors' names; do not use 
``et al.''. Papers that have not been published, even if they have been 
submitted for publication, should be cited as ``unpublished'' \cite{b4}. Papers 
that have been accepted for publication should be cited as ``in press'' \cite{b5}. 
Capitalize only the first word in a paper title, except for proper nouns and 
element symbols.

For papers published in translation journals, please give the English 
citation first, followed by the original foreign-language citation \cite{b6}.

\vspace{12pt}
\color{red}
IEEE conference templates contain guidance text for composing and formatting conference papers. Please ensure that all template text is removed from your conference paper prior to submission to the conference. Failure to remove the template text from your paper may result in your paper not being published.
-----------------------------------------------------}

\end{document}